\documentclass{wqcd03}                 % twocolumn proceedings style

\usepackage{amssymb}

\confname{QCD@Work 2003 - International Workshop on QCD, Conversano, Italy, 
14--18 June 2003}

\title{Two flavor color superconductivity and compact stars}

\author{\underline{Igor~Shovkovy}\thanks{On leave from 
Bogolyubov Institute for Theoretical Physics, 03143, Kiev, Ukraine.},
Matthias Hanauske and Mei Huang\thanks{On leave from 
Physics Department, Tsinghua University, Beijing 100084, China.} }

\address{Institut f\"{u}r Theoretische Physik,
        J.W. Goethe-Universit\"{a}t,
        D-60054 Frankurt/Main, Germany}

\begin{document}

\begin{abstract}
Baryonic matter at high density and low temperature is a color 
superconductor. In real world, this state of matter may naturally 
appear inside compact stars. A construction of a hybrid compact 
star with two flavor color superconducting quark matter in its 
interior is presented.
\end{abstract}

%% \maketitle needs to be after the author and address info and the abstract 
\maketitle

%% standard LaTeX from here on...

\section{Introduction}

At large baryon density, quantum chromodynamics (QCD) becomes a weakly
interacting theory of quarks and gluons \cite{ColPer}. Because of an 
attractive 
interaction between quarks, the ground state of such quark matter is a 
color superconductor \cite{old,bl,cs}. At asymptotic densities, quark 
matter was studied from first principles in 
Refs.~\cite{weak,weak-cfl,drqw,eff-t}. 
Unfortunately, these studies are not very reliable quantitatively at 
realistic densities that exist in nature (i.e., at densities less than 
about $10\rho_0$ where $\rho_0\approx 0.15$ fm$^{-3}$ is the normal 
nuclear density).

In general, QCD at high baryon density has a very rich phase structure.
There are many possible color superconducting phases of quark matter
made of one, two and three lightest quark flavors. Each of them is
characterised by a unique symmetry breaking pattern and by a specific
number of bosonic as well as fermionic gapless modes. 

In the rest of this paper, we are going to concentrate 
almost exclusively on matter with two quark flavors. The corresponding 
ground state of matter is the so-called two-flavor color superconductor 
(2SC). In this phase, the color gauge group SU(3)$_{c}$ is broken by 
the Anderson-Higgs mechanism down to SU(2)$_{c}$ subgroup. With the 
conventional choice of the condensate pointing in the ``blue" direction, 
one finds that the condensate consists of red up ($u_{r}$) and green 
down ($d_{g}$), as well as green up ($u_{g}$) and red down ($d_{r}$) 
quark Cooper pairs. The other two quarks ($u_{b}$ and $d_{b}$) do 
not participate in pairing. This is the conventional picture of the 
2SC phase \cite{bl,cs,weak}.

In passing, we mention that quark matter at high density may also 
contain strange quarks. This would be the case when the constituent 
medium modified mass of the strange quark is smaller than the value 
of the strange chemical potential. The current limited knowledge of 
QCD properties at finite density does not allow us to resolve the issue 
regarding the strangeness content in baryonic matter at densities 
existing inside compact stars unambiguously. As a benchmark test, 
here we study only non-strange quark matter.
 
We start our discussion by pointing that matter in the bulk of a 
compact star should be neutral with respect to electrical as well as 
color charges. Also, such matter should remain in $\beta$-equilibrium. 
Satisfying these requirements impose nontrivial relations between the 
chemical potentials of different quarks \cite{absence2sc,neutral_steiner}. 
In turn, such relations influence the pairing dynamics between quarks, 
for instance, by suppressing conventional 2SC phase and favoring the 
so-called gapless 2SC (g2SC) phase \cite{HS}. 

The condition of charge neutrality should not necessarily be satisfied 
{\em locally}. It is acceptable, for example, if matter stays in a 
mixed phase that is neutral {\em globally}, or on average 
\cite{glen92,neutral_buballa}. Below we use this idea to obtain a 
model equation of state of non-strange hybrid matter that is made 
in part of the 2SC phase \cite{original}.

\section{Quark model}
\label{quark-model}

We use the simplest SU(2) Nambu-Jona-Lasinio model of 
Ref.~\cite{huang_2sc} to describe quark matter,
\noindent
\begin{eqnarray}
{\cal L} & =
&\bar{q}(i\gamma^{\mu}\partial_{\mu}-m_0)q +
 G_S\left[(\bar{q}q)^2 + (\bar{q}i\gamma_5{\bf \vec{\tau}}q)^2\right]
\nonumber \\
 &+& G_D\left[(i \bar{q}^C  \varepsilon  \epsilon^{b} \gamma_5 q )
   (i \bar{q} \varepsilon \epsilon^{b} \gamma_5 q^C)\right],
\label{lagr}
\end{eqnarray}
where $q^C=C {\bar q}^T$ is the charge-conjugate spinor and $C=i\gamma^2
\gamma^0$ is the charge conjugation matrix. The quark field is a 
4-component Dirac spinor that carries flavor ($i=1,2$) and color 
($\alpha=1,2,3$) indices.  ${\vec \tau} =(\tau^1,\tau^2, \tau^3)$ are 
Pauli matrices in the flavor space, and $(\varepsilon)^{ik} \equiv 
\varepsilon^{ik}$, $(\epsilon^b)^{\alpha \beta} \equiv 
\epsilon^{\alpha \beta b}$ are antisymmetric tensors in
flavor and color, respectively. We introduce two independent
coupling constants in the quark-antiquark and diquark
channels, $G_S$ and $G_D$. Also we restrict ourselves only to the 
chiral limit ($m_0=0$).

The values of the parameters in the NJL model are the same as in 
Ref.~\cite{SKP}: $G_S=5.016$ GeV$^{-2}$ and the cut-off 
$\Lambda=653$ MeV. The strength of the diquark coupling $G_D$ is taken to be
proportional to the quark-antiquark coupling constant, i.e., $G_D = \eta
G_S$ with $\eta=0.75$. The choice $\eta=0.75$ corresponds to the regime
of intermediate strength of the diquark coupling \cite{HS}.

In $\beta$-equilibrium, the diagonal matrix of quark chemical potentials
is given in terms of baryon ($\mu_B\equiv 3\mu$), electrical and color 
chemical potentials as follows:
\begin{equation}
\mu_{ij, \alpha\beta}= (\mu \delta_{ij}- \mu_e Q_{ij})
\delta_{\alpha\beta} + \frac{2}{\sqrt{3}}\mu_{8} \delta_{ij}
(T_{8})_{\alpha \beta},
\end{equation}
where $Q$ and $T_8$ are the generators of U(1)$_{em}$ of electromagnetism 
and U(1)$_{8}$ subgroup of SU(3)$_{c}$ group.

In the mean field approximation, the effective potential for zero 
temperature quark matter in $\beta$-equilibrium with electrons 
takes the form \cite{HS}
\begin{center}
\begin{equation}
\Omega = \Omega_{0}-\frac{\mu_e^4}{12 \pi^2}
+\frac{m^2}{4G_S}+\frac{\Delta^2}{4G_D}
- \sum_{a} \int\frac{d^3 p}{(2\pi)^3} |E_{a}|,
\label{pot-2sc}
\end{equation}
\end{center}
where $\Omega_{0}$ is a constant added to make the pressure of the vacuum
vanishing. The sum in the last term runs over all (6 quark and 6
antiquark) quasiparticles. The dispersion relations and the degeneracy
factors of the quasiparticles read
\begin{eqnarray}
E_{ub}^{\pm} &=& E(p) \pm \mu_{ub} , \hspace{26.6mm} [\times 1]
\label{disp-ub} \\
E_{db}^{\pm} &=& E(p) \pm \mu_{db} , \hspace{26.8mm} [\times 1]
\label{disp-db}\\
E_{\Delta^{\pm}}^{\pm} &=& \sqrt{[E(p) \pm \bar{\mu}]^2
+\Delta^2} \pm  \delta \mu .\hspace{6.2mm} [\times 2]
\label{2-degenerate}
\end{eqnarray}
Here we used the following shorthand notation:
\begin{eqnarray}
E(p) &\equiv& \sqrt{{\bf p}^2+m^2}, \\
\bar{\mu} &\equiv& \frac{\mu_{ur} +\mu_{dg}}{2} =
\frac{\mu_{ug}+\mu_{dr}}{2}
=\mu-\frac{\mu_{e}}{6}+\frac{\mu_{8}}{3}, \\
\delta\mu &\equiv& \frac{\mu_{dg}-\mu_{ur}}{2}
=\frac{\mu_{dr}-\mu_{ug}}{2}
=\frac{\mu_{e}}{2}.
\end{eqnarray}
The value of the thermodynamic potential that determines the pressure, 
$\Omega_{\rm phys} =-P$, is obtained from $\Omega$ in Eq.~(\ref{pot-2sc}) 
after substituting the order parameter $\Delta$ that solves the gap 
equation, i.e., $\partial \Omega/\partial \Delta =0$.

Now, if we restrict ourselves to the case of homogeneous quark matter, 
the following two conditions of charge neutrality should be imposed:
\begin{center}
\begin{equation}
n_{8} \equiv
\frac{\partial \Omega}{\partial \mu_{8}}=0,
\qquad\qquad
n_{Q} \equiv
\frac{\partial \Omega}{\partial \mu_{e}}=0.
\label{electr-neut}
\end{equation}
\end{center}
The solution that satisfies the gap equation and both neutrality
conditions was studied in detail in Ref.~\cite{HS}. It was found that the
corresponding phase of matter is the g2SC phase. This phase has the same 
symmetry of the ground state as the conventional 2SC phase. In the 
low-energy spectrum, however, it has two additional gapless fermionic
quasiparticles. 

Our analysis shows that the neutral g2SC phase is more favorable than 
the neutral normal phase of quark matter \cite{HS}. We find, in particular,
that the pressure difference of these two phases is of order 1 MeV/fm$^{3}$.

\section{Mixed phases}

In addition to homogeneous phases, one could also study
various mixed phases of quark matter. The most promising components for 
constructing mixed phases are (i) normal phase, (ii) g2SC phase, and
(iii) ordinary 2SC phase. Note that the first two of them allow locally 
neutral phases by themselves, while the last one tends to be positively 
charged. In principle, other phases could also be taken into consideration.

Inside mixed phases, the charge neutrality is satisfied ``on average"
rather than locally. This means that different components of a mixed phase
may have non-zero densities of conserved charges, but the total charge 
of all components vanishes. In absence of {\em local} neutrality, the 
pressure of each phase could be considered as a function of baryon chemical 
potential, as well as a function of chemical potentials related to other 
conserved charges (e.g., $\mu_{e}$ and $\mu_{8}$ in the model at hand). 
This information is used as an input to obtain the most favorable mixed 
phase construction (see below).

While we intend to relax the local neutrality condition with respect to
the electrical charge, in this paper we always enforce the condition of
{\em local color} neutrality. Here we cannot fully justify this constrain. 
However, one could speculate that separation of color charges is less 
likely in mixed phases.

\begin{figure}
\hbox to\hsize{\hss
\includegraphics[bbllx=5,bblly=45,bburx=575,bbury=590,width=\hsize]{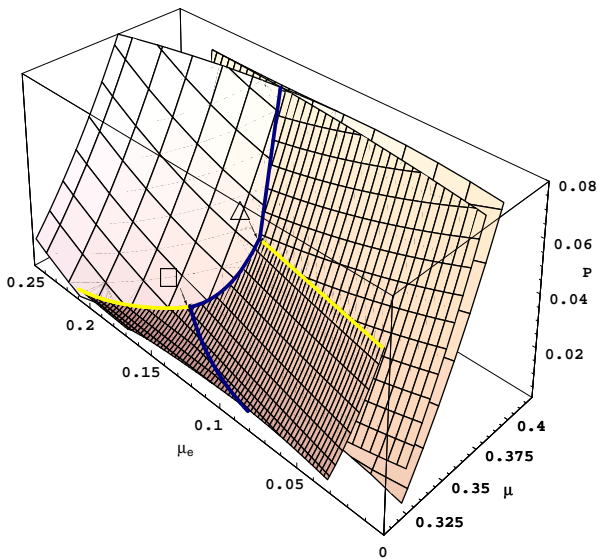}
\hss}
\caption{\label{fig:3D_HadQu}
Pressure as a function of $\mu\equiv\mu_B/3$ and $\mu_e$ for the
hadronic phase (at the bottom), for the 2SC phase (on the right 
in front) and the normal phase of quark matter (on the left). The 
dark solid line follows the neutrality line in hadronic matter, and 
two mixed phases: (i) a mixture of the hadronic and normal quark 
phases; and (ii) a mixture of the normal and 2SC quark phases.}
\end{figure}

Without going into much detail, let us give a brief introduction into 
the general method of constructing mixed phases by imposing the Gibbs 
conditions of equilibrium \cite{glen92} which are the conditions of 
mechanical and chemical equilibrium between different components. In 
regard to a mixed phase of the normal and 2SC components of quark matter, 
for example, these conditions read
\begin{eqnarray}
P^{(NQ)}(\mu,\mu_{e}) &=& P^{(2SC)}(\mu,\mu_{e}),
\label{P=P}\\
\mu &=& \mu^{(NQ)}=\mu^{(2SC)},
\label{mu=mu}\\
\mu_{e} &=& \mu^{(NQ)}_{e}=\mu^{(2SC)}_{e}.
\label{mue=mue}
\end{eqnarray}
These conditions are easy to visualize by plotting the pressure as
a function of chemical potentials ($\mu$ and $\mu_{e}$) for each
component, see Fig.~\ref{fig:3D_HadQu}. As is clear, the Gibbs 
conditions are satisfied automatically along the intersection 
lines of different pressure surfaces.

%%%%%%%%%%%%%%%%%%%%%%%%%%%%%%%%%%%%%%%%%%%%%%%%%%%%%%
%%%%%%%%%%%%%%%%%%%%%%%%%%%%%%%%%%%%%%%%%%%%%%%%%%%%%%

In a mixed phase, different components occupy different volumes of
space. To describe this quantitatively, we introduce the volume fraction
of the normal phase of quark matter: $\chi^{NQ}_{2SC}\equiv V_{NQ}/V$
(notation $\chi^{A}_{B}$ means ``volume fraction of phase A in a mixture
with phase B"). Then, the volume fraction of the 2SC phase is given by
$\chi^{2SC}_{NQ}=(1-\chi^{NQ}_{2SC})$. From the definition, it is clear
that $0\leq \chi^{NQ}_{2SC} \leq 1$.

The average electric charge density of the mixed phase is determined by
the charge densities of its components taken in the proportion of the
corresponding volume fractions. Thus,
\begin{center}
\begin{equation}
n^{(MP)}_{e} = \chi^{NQ}_{2SC} n^{(NQ)}_{e}
+(1-\chi^{NQ}_{2SC}) n^{(2SC)}_{e}.
\end{equation}
\end{center}
If the charge densities of the two components have opposite signs, the 
global charge neutrality condition, $n^{(MP)}_{e}=0$, can be imposed.
Otherwise, a neutral mixed phase could not exist. In the case of
quark matter, the charge density of the normal quark phase is negative,
while the charge density of the 2SC phase is positive along the line of
the Gibbs construction (dark solid line in Fig.~\ref{fig:3D_HadQu}).
So, a neutral mixed phase exists, and the volume fractions of its
components are
\begin{eqnarray}
\chi^{NQ}_{2SC} &=& \frac{n^{(2SC)}_{e}}{n^{(2SC)}_{e}-n^{(NQ)}_{e}}, \\
\chi^{2SC}_{NQ} &\equiv& 1-\chi^{NQ}_{2SC}=
\frac{n^{(NQ)}_{e}}{n^{(NQ)}_{e}-n^{(2SC)}_{e}}.
\end{eqnarray}
By making use of these expressions, we can calculate the energy density 
of the corresponding mixed phase,
\begin{center}
\begin{equation}
\varepsilon^{(MP)} = \chi^{NQ}_{2SC} \varepsilon^{(NQ)}
+(1-\chi^{NQ}_{2SC}) \varepsilon^{(2SC)}.
\end{equation}
\end{center}
This is essentially all that we need in order to construct the equation of
state of the mixed phase.

So far, we neglected the effects of the Coulomb forces and the
surface tension between different components of the mixed phase. In a 
real system, however, these are important. In particular, the balance
between the Coulomb forces and the surface tension determines the size
and geometry of different components of the mixed phase. In our case, 
nearly equal volume fractions of the two quark phases are likely to 
form alternating layers (slabs) \cite{geometry}. The thickness $a$
of the layers scales as $\sigma^{1/3} (n_{e}^{(2SC)}-n_{e}^{(NQ)})^{-2/3}$
where $\sigma$ is the surface tension. In the model at hand $a\approx 10$ 
fm.

The energy cost per unit volume to produce the layers scales as 
$\sigma^{2/3}(n_{e}^{(2SC)}-n_{e}^{(NQ)})^{2/3}$  \cite{geometry}. 
Therefore, the mixed phase is favorable only 
if the surface tension is not too large. Our estimates show that 
$\sigma_{max} \lesssim 20$ MeV/fm$^{2}$. For slightly larger values, 
$20 \lesssim \sigma \lesssim 50$ MeV/fm$^{2}$, the mixed phase is also 
possible, but its first appearance occurs at higher densities, 
$3\rho_0 \lesssim \rho_B \lesssim 5\rho_0$. Note that the value of 
the maximum surface tension is of the same order as in 
Refs.~\cite{geometry,interface}. In this study we assume that actual
value of the surface tension is not very large. 

The validity of the quark model is limited when the baryon density
decreases. In fact, at sufficiently low densities, quarks are confined 
inside hadrons, and it is more natural to use a hadronic model. Here
we use the chiral SU(3)$_L \times $SU(3)$_R$ model of
Ref.~\cite{papa98}.

It is expected that the hadronic and quark phases are separated by 
a first order phase transition. Then, the hadronic and quark 
phases can co-exist in a mixed phase \cite{glen92} that is constructed 
by satisfying the Gibbs conditions similar to those discussed earlier. 
Thus, we plot the hadronic 
surface of the pressure along with the quark surfaces in 
Fig.~\ref{fig:3D_HadQu}. Again, the intersection lines of different
surfaces indicate all potentially viable mixed phases. Although the 
Gibbs conditions are satisfied along all such lines, not all of them 
can produce neutral phases (e.g., there are no neutral constructions 
along the lightly shaded solid lines in Fig.~\ref{fig:3D_HadQu}).

The dark solid line in Fig.~\ref{fig:3D_HadQu} gives the full
construction that consists of three pieces. The lowest part of the 
curve (up to the point denoted by $\square$) corresponds to the neutral 
hadronic phase. Within this region matter is composed mostly of 
neutrons and a small fraction of protons and electrons. 

The mixed phase of hadronic and normal quark matter starts at the baryon
density $\rho_B\approx 1.49 \rho_0$ ($\square$-point in 
Fig.~\ref{fig:3D_HadQu}). At this point the first bubbles of deconfined 
quark matter appear in the system. At the beginning of this hadron-quark 
mixed phase, the deconfined bubbles are small but highly negatively 
charged, whereas the hadronic phase, in which the bubbles are embedded, 
is slightly positively charged. The charge neutrality condition reads
\begin{equation}
n^{(MP)}_{e} \equiv \chi^{NQ}_{H} n^{(NQ)}_{e}
+(1-\chi^{NQ}_{H}) n^{(H)}_{e}=0.
\label{eq:q_MP}
\end{equation}
where $n^{(H)}_{e}$ and $n^{(NQ)}_{e}$ are the charge densities of
hadronic and normal quark phases, respectively. This condition is
satisfied at each point along the middle part of the dark solid line
(i.e., between $\square$- and $\bigtriangleup$-points). With 
increasing density (up to $2.56 \rho_0$), the volume fraction of 
the hadronic phase decreases (down to $0.59$). However, one does not 
reach a point where the fraction of hadronic phase would vanish 
completely. Instead, at baryon density about $2.56 \rho_0$ 
($\bigtriangleup$-point in Fig.~\ref{fig:3D_HadQu}), the mixed 
phase is replaced by another mixed phase which is made of the normal 
and 2SC quark components. At this point, the positively charged 
hadronic component will be suddenly converted into a positively 
charged 2SC quark component. As a result 
of this rearrangement, the values of the baryon density and
the energy density experience small jumps: the baryon density changes
from about $2.56 \rho_0$ to $2.75 \rho_0$ and the energy density changes
from $378$ MeV$/$fm$^{3}$ to $415$ MeV$/$fm$^{3}$. After the
transition, the fractions of the 2SC and normal quark phases are
$0.53$ and $0.47$, respectively.

At higher densities, the mixture of the normal and 2SC quark 
phases is the most favorable globally neutral construction in the 
model at hand. The volume fractions of the components in this mixed 
phase stay nearly constant with increasing density.

\begin{figure}
\hbox to\hsize{\hss
\includegraphics[width=\hsize]{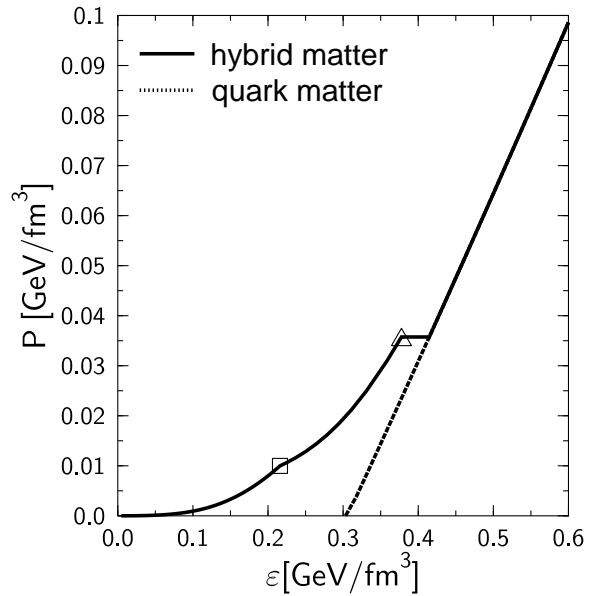}
\hss}
\caption{\label{fig:eos}
The equation of state for globally neutral hybrid matter (solid line)
and globally neutral quark matter (dashed line). The points of the
beginning of the two mixed phases are denoted by $\square$ and
$\bigtriangleup$.}
\end{figure}

The complete equation of state of hybrid baryon matter is shown in
Fig.~\ref{fig:eos} by solid line. For comparison, the equation of 
state for globally neutral quark matter is also shown. As before, 
the points that indicate the beginning of two different mixed phases 
of hybrid matter are denoted by $\square$ and $\bigtriangleup$ 
symbols.

%%%%%%%%%%%%%%%%%%%%%%%%%%%%%%%%%%%%%%%%%%%%%%%%%%%%%%
%%%%%%%%%%%%%%%%%%%%%%%%%%%%%%%%%%%%%%%%%%%%%%%%%%%%%%

\section{Star properties}
\label{star-properties}

Here we use the equation of state of hybrid matter that is shown in 
Fig.~\ref{fig:eos} to construct non-rotating compact stars. This is
done by solving the well known Tolman-Oppenheimer-Volkoff (TOV) 
equations.

\begin{figure}
\hbox to\hsize{\hss
\includegraphics[width=\hsize]{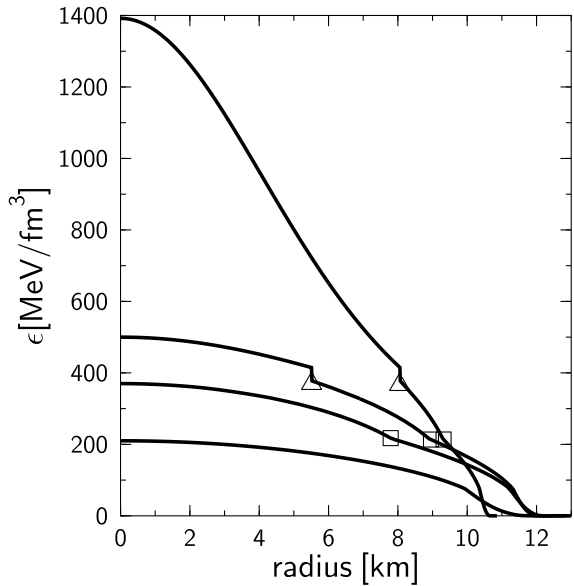}
\hss}
\caption{\label{fig:er}
Energy density profiles for hybrid stars. The locations of the interface
between the two types of mixed phases are denoted by $\bigtriangleup$,
while the locations of the boundary between the pure hadronic phase and
the hadron-quark mixed phase are denoted by $\square$.}
\end{figure}

The energy density profiles for hybrid stars with different central
energy densities are displayed in Fig.~\ref{fig:er}. The star with the
lowest central energy density in Fig.~\ref{fig:er}, $\epsilon_c=210$
MeV$/$fm$^{3}$, is composed of pure confined hadronic matter, mainly
neutrons, surrounded by a thin compact star crust consisting of leptons
and nuclei. To get the equation of state for the crust, we use the
results of Ref.~\cite{bay71a} for $\rho_B<0.001$ fm$^{-3}$ and the
results of Ref.~\cite{nege73a} for $0.001 < \rho_B < 0.08$ fm$^{-3}$.

%%%%%%%%%%%%%%%%%%%%%%%%%%%%%%%%%%%%%%%%%%%%%%%%%%%%%%
%%%%%%%%%%%%%%%%%%%%%%%%%%%%%%%%%%%%%%%%%%%%%%%%%%%%%%

The next energy density profile in Fig. \ref{fig:er} corresponds to the
central energy density $\epsilon_c=370$ MeV$/$fm$^{3}$. We see that the
corresponding star already has a rather large core (the radius is about
$8$ km) consisting of a mixture of hadronic and normal quark matter. The 
core of the star is surrounded by a layer of hadronic matter and a crust.

In the model at hand, there are no stars with central energy densities 
in the window between $378$ MeV$/$fm$^{3}$ and $415$ MeV$/$fm$^{3}$.
At $\epsilon_c>415$ MeV$/$fm$^{3}$, a quark core (made of the mixed phase
of the normal and 2SC components) forms at the center of the star. Two
examples of the corresponding energy density profiles are also shown in
Fig. \ref{fig:er}. The star with the central energy density
$\epsilon_c=500$ MeV$/$fm$^{3}$ contains a quark phase core with radius
about $6$ km. This core is separated from the layer of the hadron-quark
mixed phase by a sharp interface (the corresponding point is denoted by
$\bigtriangleup$ in Fig. \ref{fig:er}). The star with the largest 
possible mass within this model has the 
central energy density $\epsilon_c=1392$ MeV$/$fm$^{3}$. This star is
composed mainly of the quark core surrounded by a relatively thin layer
of the hadron-quark mixed phase, as well as the pure hadronic phase and
the crust on the outside.

\begin{figure}
\hbox to\hsize{\hss
\includegraphics[width=\hsize]{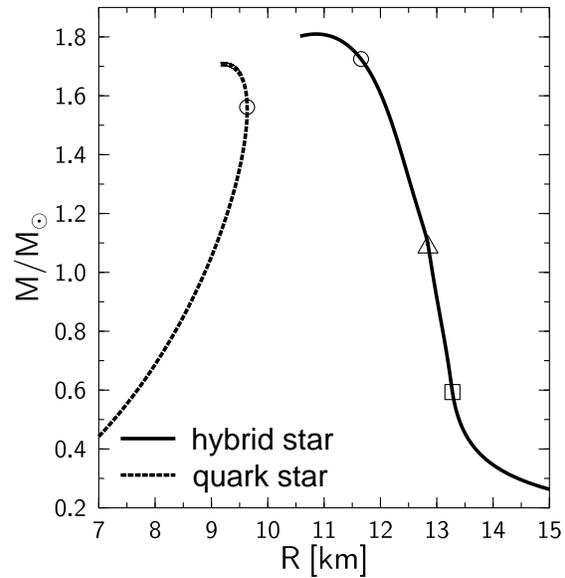}
\hss}
\caption{\label{fig:mr}
The mass-radius relations for hybrid stars (solid line) and quark stars
(dashed line).}
\end{figure}

The results for the mass-radius relation of hybrid and quark stars 
are shown in Fig.~\ref{fig:mr}. As one would expect, the pure quark 
stars have much smaller radii and the value of their maximum mass is 
slightly smaller. The difference between hybrid and pure quark stars is
mostly due to the low density part of the equation of state. This is also
evident from the qualitative difference in the dependence of the radius
as a function of mass for the hybrid and quark stars with low masses. The
corresponding hybrid stars are large because of a sizable low density
hadronic layer, while the quark stars are small because they have no such
layers.

%%%%%%%%%%%%%%%%%%%%%%%%%%%%%%%%%%%%%%%%%%%%%%%%%%%%%%
%%%%%%%%%%%%%%%%%%%%%%%%%%%%%%%%%%%%%%%%%%%%%%%%%%%%%%

Our results for pure quark stars are comparable to those in Refs.
\cite{Blaschke_2sc,Ruester,super-dense,compact_CFL}. Also, the maximum
masses and the corresponding radii of the hybrid stars obtained here are
similar to those of the strange hybrid stars of Ref. \cite{super-dense},
provided the strange quark mass is not very small ($m_s\gtrsim 300$ MeV)
and the superconducting gap is not too large ($\Delta \lesssim 50$ MeV). 
At small values of the strange quark mass and/or large values of the
superconducting gap, the strange hybrid stars tend to have smaller
maximum masses and smaller radii \cite{super-dense}.

\section{Summary and outlook}
\label{summary}

Here we constructed a realistic equation of state of non-strange
baryonic matter that is globally neutral and satisfies the condition 
of $\beta$-equilibrium. This equation of state might be valid up to
densities of about $8\rho_0$
in the most optimistic scenario. In our construction, matter at low
density ($\rho_B\lesssim 1.49 \rho_0$) is mostly made of neutrons with traces
of protons and electrons. At intermediate densities ($1.49 \rho_0 \lesssim
\rho_B \lesssim 2.56 \rho_0$), homogeneous hadronic matter is replaced by the
mixed phase of positively charged hadronic matter and deconfined bubbles
of negatively charged normal quark matter. The volume fraction of normal
quark matter gradually grows with increasing density. Before the volume
fractions of two phases become equal, the mixed phase undergoes a
rearrangement in which the hadronic component of matter turns into the
two-flavor color superconductor.  At higher densities ($\rho_B \gtrsim 2.75
\rho_0$), only the quark mixed phase exists. This latter is composed of
about equal fractions of normal and 2SC quark matter.

Previously, it was argued that 2SC quark matter could not appear in
compact stars when the charge neutrality condition is imposed locally
\cite{absence2sc}. The main reason for this is the strong preference of the 
2SC phase to remain positively charged. Our study shows, however, that the 
positively charged 2SC quark component appears naturally in a quark mixed 
phase at densities around $3\rho_0$. The other component of the globally 
neutral mixed phase is negatively charged normal quark matter. The 
corresponding construction turns out to be rather stable. In particular, 
we observe that the volume fractions of the two quark components remain 
approximately the same with changing the baryon density in a wide range. 

By making use of the equation of state of hybrid matter, we construct
non-rotating compact stars. We find that the largest mass hybrid 
star has the radius $10.86$ km, the mass $1.81 M_{\odot}$ and the 
central baryon density $7.58 \rho_0$. This star has a large ($8$ km)
quark core, and a relatively thin outer layers of hadronic matter and 
a crust. One could speculate that the appearance of a large core is 
connected with the fact that the quark phase starts to develop at 
relatively low densities, $\rho_B\approx 2.75\rho_0$, in the model 
used.

In the end, we would like to mention that performing a systematic 
study of the model dependence of hybrid matter constructions with
various color superconducting phases would be an important task for
future. Such a study will be crucial for resolving some apparent 
differences between the results currently existing in the literature 
regarding non-strange as well as strange quark matter 
\cite{original,Blaschke_2sc,Ruester,super-dense,compact_CFL}.

\section*{Acknowledgements}
I.S. would like to thank the organizers of the Workshop QCD@Work 2003 
for their kind hospitality during the meeting, as well as for a local 
financial support. The work I.S. of was supported in part by 
Gesellschaft f\"{u}r Schwerionenforschung (GSI) and by Bundesministerium 
f\"{u}r Bildung und Forschung (BMBF).

\end{document}